\documentclass[nofootinbib,twocolumn,showpacs,preprintnumbers,pre,aps]{revtex4-1}

\usepackage{graphicx}
\usepackage{bm}
\usepackage{amsmath}
\usepackage{amssymb}
\usepackage{color}

\begin{document}

\title{Thermally Driven Elastic Micromachines}%

\author{Yuto Hosaka}

\affiliation{
Department of Chemistry, Graduate School of Science and Engineering,
Tokyo Metropolitan University, Tokyo 192-0397, Japan}

\author{Kento Yasuda}

\affiliation{
Department of Chemistry, Graduate School of Science and Engineering,
Tokyo Metropolitan University, Tokyo 192-0397, Japan}

\author{Isamu Sou}

\affiliation{
Department of Chemistry, Graduate School of Science and Engineering,
Tokyo Metropolitan University, Tokyo 192-0397, Japan}

\author{Ryuichi Okamoto}

\affiliation{
Research Institute for Interdisciplinary Science, Okayama University,
Okayama 700-8530, Japan}

\author{Shigeyuki Komura}\email{komura@tmu.ac.jp}

\affiliation{
Department of Chemistry, Graduate School of Science and Engineering,
Tokyo Metropolitan University, Tokyo 192-0397, Japan}

\date{\today}

\begin{abstract}
We discuss the directional motion of an elastic three-sphere micromachine in which 
the spheres are in equilibrium with independent heat baths having different temperatures.
Even in the absence of prescribed motion of springs, such a micromachine can 
gain net motion purely because of thermal fluctuations. 
A relation connecting the average velocity and the temperatures of the spheres 
is analytically obtained.
This velocity can also be expressed in terms of the average heat flows in the steady state.
Our model suggests a new mechanism for the locomotion of micromachines in nonequilibrium 
biological systems.
\end{abstract}

\maketitle

Microswimmers are tiny machines that swim in a fluid, such as sperm cells or motile bacteria, 
and are expected to be applied to microfluidics and microsystems~\cite{Lauga092}.
By transforming chemical energy into mechanical work, these objects change their 
shape and move in viscous environments.
Over the length scale of micromachines, the fluid forces acting on them are governed by 
viscous dissipation.
According to Purcell's scallop theorem~\cite{Purcell77}, time-reversal body motion
cannot be used for locomotion in a Newtonian fluid. 
As one of the simplest models exhibiting broken time-reversal symmetry, 
Najafi and Golestanian proposed a three-sphere swimmer~\cite{Golestanian04,Golestanian08}, 
in which three in-line spheres are linked by two arms of varying length.
Recently, Pande \textit{et al.}\ and the present authors independently proposed a 
generalized three-sphere microswimmer in which the spheres are connected by two elastic 
springs~\cite{Pande17,Yasuda17-2}.

In the previous three-sphere microswimmer models, either the arm lengths or the natural 
lengths of the springs were assumed to undergo prescribed cyclic 
motions~\cite{Golestanian04,Golestanian08,Pande17,Yasuda17-2}.
Such active motions can lead to net locomotion if the swimming strokes are nonreciprocal.
From a practical point of view, however, it is not a simple task to implement these motions 
at micron length scales. 
Another approach for extending the Najafi--Golestanian model is to consider the arm motions 
as occurring stochastically~\cite{GolestanianAjdari08,Golestanian10,Sakaue10,Huang12}.
Although proteins or enzymes are naturally designed to include such sophisticated 
molecular mechanisms, it is still a substantial challenge to construct them artificially.
It should also be noted that thermal agitations due to surrounding fluids become 
more significant at these small scales.

In this letter, by using an elastic three-sphere micromachine~\cite{Pande17,Yasuda17-2}, 
we suggest a new mechanism for locomotion that is purely induced by thermal 
fluctuations. 
To highlight this effect, we do not consider any prescribed motion of the natural 
lengths~\cite{Yasuda17-2}.
On the other hand, the key assumption in our model is that the three spheres are in 
equilibrium with independent heat baths having different temperatures.
In this case, heat transfer occurs from a hotter sphere to a colder one, driving the 
whole system out of equilibrium. 
We show that a combination of heat transfer and hydrodynamic interactions 
among the spheres can lead to directional locomotion in the steady state. 
We analytically obtain the expression for the average velocity in terms of the  
sphere temperatures.
Our finding is further confirmed by numerical simulations.
Since our model has a similarity to a class of thermal ratchet models that have 
been intensively studied before~\cite{Sekimoto97,Sekimoto98,SekimotoBook}, 
the suggested mechanism is relevant to nonequilibrium dynamics of proteins and 
enzymes in biological systems.

\begin{figure}[b]
\begin{center}
\includegraphics[scale=0.35]{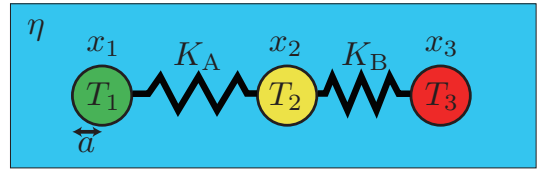}
\end{center}
\caption{
(Color online)　
Thermally driven elastic three-sphere micromachine in a fluid of viscosity $\eta$.
Three spheres of radius $a$ are connected by two harmonic springs with elastic 
constants $K_{\rm A}$ and $K_{\rm B}$.
The time-dependent positions of the spheres are denoted by $x_i(t)$ ($i=1, 2, 3$) 
in a one-dimensional coordinate system. 
Importantly, the three spheres are in equilibrium with independent heat baths at 
temperatures $T_i$.
}
\label{model}
\end{figure}

As schematically shown in Fig.~\ref{model}, we consider a three-sphere micromachine and 
take into account the elasticity in the internal spring motions~\cite{Pande17,Yasuda17-2}.
This model consists of three hard spheres of radius $a$ connected by two 
harmonic springs A and B with spring constants $K_{\rm A}$ and $K_{\rm B}$, respectively.
The natural length of the springs, $\ell$, is assumed to be constant.
The total energy is given by 
\begin{align}
E = \frac{K_{\rm A}}{2}(x_2 - x_1 - \ell)^2 + 
\frac{K_{\rm B}}{2}(x_3 - x_2 - \ell)^2,
\end{align}
where $x_i(t)$ ($i=1, 2, 3$) are the positions of the three spheres in a one-dimensional 
coordinate system and we  assume $x_1<x_2<x_3$ without loss of generality.
Owing to the hydrodynamic interactions, each sphere exerts a force on the viscous fluid 
of shear viscosity $\eta$ and experiences an opposite force from it.
In general, the surrounding medium can be viscoelastic~\cite{Yasuda17}, but such an 
effect is not included in this letter.

We consider a situation in which the three spheres are in equilibrium with independent 
heat baths at temperatures $T_i$.
When these temperatures are different, the system is driven out of equilibrium
because a heat flux is generated from a hotter sphere to a colder one.
Denoting the velocity of each sphere by $\dot x_i$, we can write the equations of motion 
of the three spheres  as 
\begin{align}
\dot x_1& =\frac{K_{\rm A}}{6\pi\eta a}(x_2-x_1-\ell)
-\frac{K_{\rm A}}{4\pi\eta}\frac{(x_2-x_1-\ell)}{x_2-x_1}
\nonumber \\
& +\frac{K_{\rm B}}{4\pi\eta}\frac{(x_3-x_2-\ell)}{x_2-x_1}
-\frac{K_{\rm B}}{4\pi\eta}\frac{(x_3-x_2-\ell)}{x_3-x_1} +\xi_1,  
\label{V1} \\
\dot x_2 &=\frac{K_{\rm A}}{4\pi\eta}\frac{(x_2-x_1-\ell)}{x_2-x_1}
-\frac{K_{\rm A}}{6\pi\eta a}(x_2-x_1-\ell)
\nonumber \\
&+\frac{K_{\rm B}}{6\pi\eta a}(x_3-x_2-\ell)
-\frac{K_{\rm B}}{4\pi\eta}\frac{(x_3-x_2-\ell)}{x_3-x_2} + \xi_2,  
\label{V2} \\
\dot x_3 &=\frac{K_{\rm A}}{4\pi\eta}\frac{(x_2-x_1-\ell)}{x_3-x_1}
-\frac{K_{\rm A}}{4\pi\eta}\frac{(x_2-x_1-\ell)}{x_3-x_2}
\nonumber \\
& +\frac{K_{\rm B}}{4\pi\eta}\frac{(x_3-x_2-\ell)}{x_3-x_2}
-\frac{K_{\rm B}}{6\pi\eta a}(x_3-x_2-\ell) + \xi_3,
\label{V3} 
\end{align}
where we have used the Stokes' law for a sphere and the Oseen tensor in a three-dimensional 
viscous fluid.

Furthermore, the white-noise sources $\xi_i(t)$ have zero mean, 
$\langle \xi_i(t) \rangle =0$, and their correlations satisfy~\cite{DoiBook}
\begin{align}
\langle \xi_i(t)\xi_j(t')\rangle =2 D_{ij} \delta(t-t'), 
\label{FDTreal}
\end{align}
where $D_{ij}$ is the mutual diffusion coefficient.
When $i=j$, $D_{ii}$ is simply given by the Stokes--Einstein relation, i.e.,
\begin{align}
D_{ii}=\frac{k_{\rm B}T_i}{6 \pi \eta a},
\label{SE}
\end{align}
where $k_{\rm B}$ is the Boltzmann constant.
When $i \neq j$, on the other hand, we assume the following general relation:
\begin{align}
D_{ij}=\frac{k_{\rm B}\Theta(T_i, T_j)}{4 \pi \eta \vert x_i -x_j \vert},
\label{Dij}
\end{align}
where $\Theta(T_i, T_j)$ is a function of $T_i$ and $T_j$. 
For example, the relevant effective temperature can be the mobility-weighted 
average~\cite{Grosberg15},  which in the present case is given by  
$\Theta(T_i, T_j)=(T_i + T_j)/2$
because all the spheres have the same size.
However, its explicit functional form is not needed here, and it only needs to
satisfy an appropriate fluctuation dissipation theorem in thermal equilibrium, i.e., 
$T_i=T_j$.
This is because we only consider the limit of $a \ll \ell$ in the present study.

It is convenient to introduce a characteristic time scale given by 
$\tau=6\pi\eta a/K_{\rm A}$. 
We further define the ratio between the two spring constants as 
$\lambda = K_{\rm B}/K_{\rm A}$. 
We shall denote the two spring extensions by
\begin{align}
u_{\rm A}(t) =x_2-x_1-\ell,~~~~~
u_{\rm B}(t)=x_3-x_2-\ell.
\label{uAuB}
\end{align}
Notice that these quantities are related to the sphere velocities in Eqs.~(\ref{V1})--(\ref{V3}) as
$\dot u_{\rm A} = \dot x_2- \dot x_1$ and $\dot u_{\rm B} = \dot x_3- \dot x_2$, respectively. 
In the following analysis, we generally assume that $u_{\rm A}, u_{\rm B} \ll \ell$ 
as well as $a \ll \ell$, and focus only on the leading-order contribution.

To present the essential outcome of the model, we first consider the 
simplest symmetric case, i.e., $K_{\rm A}=K_{\rm B}$ ($\lambda=1)$.
We introduce the bilateral Fourier transform for any function $f(t)$ as 
$f(\omega) = \int_{-\infty}^{\infty} dt\, f(t) e^{-i\omega t}$  
and the inverse transform as 
$f(t) = \int_{-\infty}^{\infty} (d\omega/2\pi) \, f(\omega) e^{i\omega t}$.
Solving the time derivative of Eq.~(\ref{uAuB}) with the aid of Eqs.~(\ref{V1})--(\ref{V3}) in the 
frequency domain, we obtain 
\begin{align}
u_{\rm A}(\omega)&\approx \frac{(2+i\omega\tau) \xi_1(\omega)
-(1+i\omega\tau)\xi_2(\omega)-\xi_3(\omega)}{-3-4i\omega\tau+(\omega\tau)^2}\tau
\nonumber \\
&+\mathcal{O}(a),
\label{u1}
\end{align}
\begin{align}
u_{\rm B}(\omega)&\approx \frac{\xi_1(\omega)+(1+i\omega\tau) \xi_2(\omega)
-(2+i\omega\tau)\xi_3(\omega)}{-3-4i\omega\tau+(\omega\tau)^2}\tau
\nonumber \\
&+\mathcal{O}(a),
\label{u2}
\end{align}
where $\mathcal{O}(a)$ indicates the terms of the order of $a$.

The velocity of a three-sphere micromachine is generally given by 
$V(t) = (\dot x_1+\dot x_2+\dot x_3)/3$, which now becomes
\begin{align}
V(t) &\approx \frac{a}{4\ell\tau} \left( u_{\rm B}-u_{\rm A}
+\frac{u_{\rm B}^2}{2 \ell}-\frac{u_{\rm A}^2}{2\ell}  \right)
\nonumber \\
& +\frac{1}{3}(\xi_1+\xi_2+\xi_3) + \mathcal{O}(u_{\rm A}^3,u_{\rm B}^3).
\label{meanV}
\end{align}
By taking its statistical average, we further obtain 
\begin{align}
\langle V (t) \rangle \approx \frac{a}{8\ell^2\tau}\langle u_{\rm B}^2(t) 
-u_{\rm A}^2(t) \rangle + \mathcal{O}(a^2,u_{\rm A}^3,u_{\rm B}^3), 
\label{aveVreal}
\end{align} 
where we have used $\langle u_{\rm A}(t) \rangle = \langle u_{\rm B}(t) \rangle =0$ 
in the lowest-order of $a$. 
In the Fourier domain, Eq.~(\ref{aveVreal}) can be written in terms of convolution as 
\begin{align}
\langle V (\omega)\rangle &\approx \frac{a}{8\ell^2\tau}\int_{-\infty}^{\infty} \frac{d\omega'}{2\pi}\, 
\langle u_{\rm B}(\omega-\omega') u_{\rm B}(\omega')
\nonumber \\
& - u_{\rm A}(\omega-\omega') u_{\rm A}(\omega')\rangle
+ \mathcal{O}(a^2,u_{\rm A}^3,u_{\rm B}^3). 
\label{aveVomega}
\end{align}

Next, we substitute Eqs.~(\ref{u1}) and (\ref{u2}) into Eq.~(\ref{aveVomega}) and 
use the relation 
$\langle \xi_i(\omega)\xi_j(\omega')\rangle=(2\pi)(2D_{ij})\delta(\omega+\omega')$,
as directly obtained from Eq.~(\ref{FDTreal}). 
After some calculation, we have
\begin{align}
\langle V (\omega) \rangle
=\frac{a\tau}{4\ell^2}(D_{33}-D_{11})\frac{2\pi i}{4i\tau-\omega\tau^2}\delta(\omega).
\label{Vomega}
\end{align}
Notice that the cross correlations for $i \neq j$ can be neglected here  
because these are higher-order contributions of $\mathcal{O}(a^2)$. 
Transforming back to the time domain, we obtain the average velocity as 
\begin{align}
\langle V\rangle=\frac{a k_{\rm B}(T_3-T_1)}{16\tau K_{\rm A}\ell^2}=
\frac{k_{\rm B}(T_3-T_1)}{96\pi\eta\ell^2},
\label{Vsymmetric}
\end{align}
where we have used Eq.~(\ref{SE}).

The above expression is an important result of this letter and deserves further discussion.
The average velocity is proportional to the temperature difference $T_3-T_1$.
Since we have assumed $x_1<x_2<x_3$, the swimming direction is from a colder 
sphere to a hotter one, i.e., $\langle V \rangle >0$ when $T_3>T_1$
and vice versa. 
It is also remarkable that Eq.~(\ref{Vsymmetric}) does not depend on the temperature 
$T_2$ of the middle sphere.  
Hence $\langle V \rangle =0$ when $T_1=T_3$ even though $T_1$ and $T_3$ can be 
different from $T_2$. 
However, the presence of the middle sphere is essential for directional locomotion
because the hydrodynamic interactions among the three spheres are responsible for it.
Notice that a two-sphere micromachine cannot move even if the temperatures are different.  
This is because, if we keep only the first two terms in Eqs.~(\ref{V1}) and (\ref{V2}) plus 
the noise terms $\xi_1$ and $\xi_2$ for the two spheres, we can immediately see that 
$\langle \dot x_1 + \dot x_2 \rangle = \langle \xi_1 \rangle +  \langle \xi_2 \rangle =0$.

Having discussed the simplest symmetric case, we now present the result for general 
asymmetric cases when $K_{\rm A}\neq K_{\rm B}$ ($\lambda \neq 1$).
By repeating the same calculation as before, the two spring extensions in Eq.~(\ref{uAuB}) 
now become
\begin{align}
u_{\rm A}(\omega)&\approx\frac{(2\lambda+i\omega\tau) \xi_1(\omega)
-(\lambda+i\omega\tau) \xi_2(\omega)-\lambda \xi_3(\omega)}
{-3\lambda -2(1+\lambda)i\omega\tau+(\omega\tau)^2}\tau
\nonumber \\
& +\mathcal{O}(a), 
\label{uAgeneral} \\
u_{\rm B}(\omega)&\approx \frac{\xi_1(\omega)
+(1+i\omega\tau) \xi_2(\omega)-(2+i\omega\tau) \xi_3(\omega)}
{-3\lambda-2(1+\lambda)i\omega\tau+(\omega\tau)^2}\tau
\nonumber \\
& +\mathcal{O}(a).
\label{uBgeneral}
\end{align}
Then, the average velocity is 
\begin{align}
\langle  V(t) \rangle & \approx \frac{a}{8\ell^2 \tau}\langle \lambda u_{\rm B}^2(t) -
u_{\rm A}^2(t) + 3(1-\lambda) u_{\rm A}(t) u_{\rm B}(t) \rangle
\nonumber \\ 
& +\mathcal{O}(a^2,u_{\rm A}^3,u_{\rm B}^3).
\label{aveVrealgen}
\end{align}
The substitution of Eqs.~(\ref{uAgeneral}) and (\ref{uBgeneral}) into the Fourier-transformed expression 
of Eq.~(\ref{aveVrealgen}) yields the average velocity $\langle V(\omega) \rangle$ similar 
to Eq.~(\ref{Vomega}).
By performing the inverse Fourier transform, we finally obtain the general expression for the 
average velocity:
\begin{align}
\langle V\rangle & =\frac{k_{\rm B}}{144\pi\eta\ell^2(1+\lambda)}
[(2 -5 \lambda)T_1 
\nonumber \\
& -(7 - 7 \lambda)T_2+(5-2\lambda)T_3].
\label{Vgeneral}
\end{align}

When $\lambda=1$, Eq.~(\ref{Vgeneral}) reduces to Eq.~(\ref{Vsymmetric}), as expected. 
When the three temperatures are identical, i.e., $T_1=T_2=T_3$, one can also show that the 
velocity vanishes, $\langle V\rangle =0$.
This indicates that an elastic three-sphere micromachine can attain a finite velocity owing 
to the temperature difference among the spheres, rather than its structural asymmetry.

\begin{figure}[tbh]
\begin{center}
\includegraphics[scale=0.35]{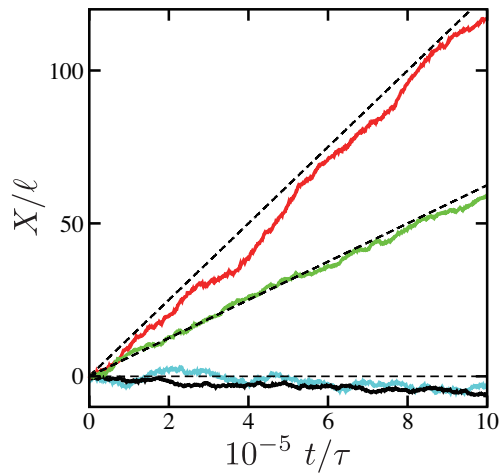}
\end{center}
\caption{
(Color online)　 Simulations of scaled center-of-mass position $X/\ell$ of an 
elastic micromachine as a function of scaled time $t/\tau$
when $a/\ell=0.1$. 
The strength of thermal noise is $S_i=[2k_{\rm B}T_i/(K_{\rm A}\ell^2)]^{1/2}$.
Case (i): thermally asymmetric nonequilibrium three-sphere micromachine with $S_1=S_2=0$ 
and $S_3=0.2$ (red) or $S_3=0.141$ (green). 
Case (ii): thermally symmetric equilibrium three-sphere micromachine with $S_1=S_2=S_3=0.067$ (black).
Case (iii): thermally asymmetric nonequilibrium two-sphere micromachine with $S_1=0$ and $S_2=0.133$ (cyan).
The dashed lines are the plots of Eq.~(\ref{Vsymmetric}) with the respective parameters.}
\label{simulation}
\end{figure}

To confirm our analytical prediction, we performed numerical simulations 
of the coupled stochastic equations in Eqs.~(\ref{V1})--(\ref{V3}) when  $\lambda=1$.
The equations can be discretized according to Storatonovich interpretation~\cite{SekimotoBook}.
Then, the strength of thermal noise acting on each sphere is determined by a dimensionless 
parameter $S_i=[2k_{\rm B}T_i/(K_{\rm A}\ell^2)]^{1/2}$.
We performed simulations for 
(i) thermally asymmetric nonequilibrium cases in which $S_1=S_2=0$ and $S_3=0.2$ or 
$S_3=0.141$, and 
(ii) a thermally symmetric equilibrium case in which $S_1=S_2=S_3=0.067$.
For comparison, we also performed a simulation for (iii) a thermally asymmetric nonequilibrium 
two-sphere case in which $S_1=0$ and $S_2=0.133$ (sphere 3 does not exist).
The average over 100 independent runs has been taken for each case.

In Fig.~\ref{simulation}, we plot the obtained center-of-mass position 
$X(t)=(x_1+x_2+x_3)/3$ 
as a function of time $t$. 
For case (i), we clearly see that a micromachine migrates towards 
the positive direction with well-defined finite velocities.
The dashed lines correspond to the analytical result in Eq.~(\ref{Vsymmetric}), 
which is in good agreement with the numerical simulations.
However, a micromachine cannot gain any net displacement for cases (ii) and (iii).
Our simulation result clearly demonstrates that a three-sphere micromachine 
can acquire directional motion because of thermal fluctuations only when the 
three spheres have different temperatures.

In the above numerical simulations, a three-sphere micromachine undergoes not only  
ballistic motion but also diffusive motion due to the presence of thermal fluctuations.
The crossover time separating these two different regimes can be roughly estimated 
by the condition $2 D t^{\ast} \approx \langle V \rangle^2 t^{\ast 2}$, where the total 
diffusion coefficient is approximately given by  
$D \approx k_{\rm B} \overline{T}/(18 \pi \eta a)$ with $\overline{T}=(T_1+T_2+T_3)/3$.
By denoting the temperature difference in Eq.~(\ref{Vsymmetric}) as $\Delta T=T_3-T_1$, 
the crossover time is roughly obtained as   
$t^* \approx D/\langle V\rangle^2 \approx \eta\ell^4 \overline{T}/[ak_{\rm B}(\Delta T)^2]$. 
When $S_1=S_2=0$ and $S_3=0.2$, for example, we estimate $t^* \approx 300\tau$,
which is much smaller than the total simulation time in Fig.~\ref{simulation}.

Next, we argue that the analytically obtained velocity in Eqs.~(\ref{Vsymmetric}) 
or (\ref{Vgeneral}) can be related to the ensemble average of heat flows 
in the steady state. 
Within the framework of ``stochastic energetics" proposed by Sekimoto~\cite{SekimotoBook},
the heat gained by the $i$-th sphere per unit time is expressed as
\begin{align}
\frac{dQ_i}{dt}=6\pi \eta a (-\dot x_i+\xi_i) \dot x_i,
\label{dQdt}
\end{align}
where $\dot x_i$ and $\xi_i$ are given by Eqs.~(\ref{V1})--(\ref{V3}).
In the calculation of average heat flows, we also consider terms up to the leading-order 
contribution of $a/\ell$.
For example, only the first term on the r.h.s.\ of Eq.~(\ref{V1}), 
$K_{\rm A}u_{\rm A}/(6\pi\eta a)$, and the noise term, $\xi_1$, are taken into 
account when we eliminate $\dot x_1$ in Eq.~(\ref{dQdt}).
We further use the statistical properties of  
quantities such as $\langle u_{\rm A}^2 \rangle$ and $\langle u_{\rm A} \xi_1 \rangle$, 
which can be estimated according to Eqs.~(\ref{uAgeneral}) and (\ref{FDTreal}).

Then, the lowest-order average heat flows are obtained as  
\begin{align}
\left\langle \frac{dQ_1}{dt}\right\rangle_0 & =  
\frac{k_{\rm B}}{6(1+\lambda)\tau}
[(3 + 2 \lambda)T_1 -(3+\lambda)T_2 -\lambda T_3],
\label{Q1}
\\
\left\langle \frac{dQ_2}{dt}\right\rangle_0 & =  
\frac{k_{\rm B}}{6(1+\lambda)\tau}
[-(3 +\lambda) T_1+(3 +2\lambda+3\lambda^2)T_2 
\nonumber \\
&-(\lambda+3\lambda^2)T_3],
\label{Q2}
\\
\left\langle \frac{dQ_3}{dt}\right\rangle_0 & =  
\frac{k_{\rm B}}{6(1+\lambda)\tau}
[-\lambda T_1 -(\lambda+3 \lambda^2)T_2
\nonumber \\ 
& +(2 \lambda+3 \lambda^2)T_3],
\label{Q3}
\end{align}
which all vanish when $T_1=T_2=T_3$.
It is also remarkable that the above lowest-order heat flows satisfy~\cite{Sekimoto98,SekimotoBook}
\begin{align}
\left\langle \frac{dQ_1}{dt}\right\rangle_0+\left\langle \frac{dQ_2}{dt}\right\rangle_0
+\left\langle \frac{dQ_3}{dt}\right\rangle_0=0.
\label{zerosum}
\end{align}
Assuming a linear relation between the velocity in Eq.~(\ref{Vgeneral}) and 
the heat flows in Eqs.~(\ref{Q1})-(\ref{Q3}), we obtain an alternative expression 
for the velocity:
\begin{align}
\langle V \rangle =\frac{a}{8K_{\rm A}\ell^2}
\left[ 
\frac{3-5\lambda}{1+\lambda} \left\langle \frac{dQ_1}{dt}\right\rangle_0
+ \frac{5-3\lambda}{\lambda(1+\lambda)}\left\langle \frac{dQ_3}{dt}\right\rangle_0
\right].
\label{vel-heat}
\end{align}
For the symmetric case of $\lambda=1$ corresponding to Eq.~(\ref{Vsymmetric}), 
the above expression reduces to
\begin{align}
\langle V \rangle=\frac{a}{8K_{\rm A}\ell^2}\left[\left\langle \frac{dQ_3}{dt}\right\rangle_0-
\left\langle \frac{dQ_1}{dt}\right\rangle_0 \right].
\label{vel-heat-sym}
\end{align}
This relation indicates that the average velocity is determined by the net heat flow between 
spheres 1 and 3.

Finally, we briefly comment on previous relevant works. 
Using coupled Langevin equations, Dunkel and Zaid investigated the interplay between 
the diffusive and self-driven behaviors of an elastic three-sphere swimmer~\cite{Dunkel09}. 
In this work, however, the temperature of the system was assumed to be uniform.
In addition, hydrodynamic simulations of a self-thermophoretic Janus particle were 
reported in Ref.~\cite{Yang14} to reproduce the experimental result~\cite{Jiang10}.
Again, our model differs from this model because thermal fluctuations of internal degrees of 
freedom cause the locomotion of an elastic micromachine.
We also note from Eq.~(\ref{Vgeneral}) that $\langle V \rangle \neq 0$ for symmetric 
temperatures $T_1=T_3 \neq T_2$ as long as the structural asymmetry exists.

In summary, we have shown that an elastic three-sphere micromachine in a viscous 
fluid can acquire directional motion because of thermal fluctuations when the spheres have 
different temperatures.
We have obtained an expression for the average velocity that is related to the temperatures 
and average heat flows.
Such a mechanism for the locomotion of micromachines is expected to play important roles 
in nonequilibrium biological systems.
In the future, we shall generalize our calculation to the case in which the spheres have 
different sizes~\cite{Golestanian08}.
In such a calculation, one needs to take into account higher-order contributions in $a$ 
and $u_{\rm A}$, $u_{\rm B}$.
It would be interesting to investigate how these nonlinear contributions affect the nonequilibrium 
dynamics of thermally driven micromachines.

S.K.\ and R.O.\ acknowledge support from a Grant-in-Aid for Scientific Research on
Innovative Areas ``\textit{Fluctuation and Structure}" (Grant No.\ 25103010) from the Ministry
of Education, Culture, Sports, Science, and Technology of Japan and from 
a Grant-in-Aid for Scientific Research (C) (Grant No.\ 15K05250)
from the Japan Society for the Promotion of Science (JSPS).


\end{document}